\documentclass[useAMS,usenatbib]{mn2e}
\usepackage{bethmacros}
\usepackage{epsfig}
\DeclareGraphicsExtensions{.eps, .jpg}
\newcommand{\beq}{\begin {equation}}
\newcommand{\eeq}{\end{equation}}

\newcommand{\bt}{\begin{table}}
\newcommand{\et}{\end{table}}

\newcommand{\beqa}{\begin{IEEEeqnarray}{rCl}}
\newcommand{\eeqa}{\end{IEEEeqnarray}}

\def\citeapos#1{\citeauthor{#1}'s (\citeyear{#1})}

\def \msun {M$_{\sun}$}

\title[Diverse satellite luminosity functions]{The luminosity function of diverse satellite galaxy systems}

\author[Nickerson et al.]{
{S. Nickerson$^{1}$, G. Stinson$^{2}$, H. M. P. Couchman$^{3}$, J. Bailin$^{4}$,  J. Wadsley$^{3}$}
\vspace*{6pt}\\
$^1$Institute for Theoretical Physics, University of Z\"{u}rich, Switzerland\\
$^2$Max-Planck-Institut f\"{u}r Astronomie, Heidelberg, Germany\\
$^3$Physics and Astronomy, McMaster University, Hamilton, Canada\\
$^4$Department of Physics and Astronomy, University of Alabama, Tuscaloosa, AL, United States}

\begin{document}
\maketitle
\label{firstpage}

\begin{abstract}
The high-resolution, SPH galaxies of the McMaster Unbiased Galaxy Survey (MUGS) are used to examine the satellite systems of sixteen model host galaxies. Each galaxy has a different mass, angular momentum and merger history that yield a rich set of satellite luminosity functions. With new observations of distant satellite systems, we can compare these luminosity functions to satellite systems beyond the Local Group. We find that the luminosity functions of our simulations compare well to observations when the luminosity functions are scaled according to host mass. We use the recently-found relationship between dwarf satellites and host mass in distant satellite systems \citep{Trentham2009} to normalize a theoretical, complete luminosity function for the Milky Way \citep{Koposov2008}. The luminosity function of satellites, expressed as a function of the host mass, is given by $dN/dM_V=3.5M_{host}^{0.91}\times10^{0.1M_V-10.2}$, where mass is given in $\Msun$. The mass of a host galaxy can be used to predict the number of dwarf satellites and even when considering spiral and elliptical hosts separately this relation holds.
\end{abstract}


\section{Introduction}
\label{sec:intro}

Dwarf galaxies are the most common objects in the Universe today \citep{Marzke1997}. Their numbers provide insight into many aspects of the evolution of the Universe from the composition of dark matter to how star formation affects galaxy formation.  Dwarf galaxies are however difficult to observe because of their low surface brightness and a detailed census of dwarf galaxies on large scales remains a daunting challenge.  Only for the most nearby systems is a complete counting of dwarf galaxies possible.  Such accounting leads to the well-known discrepancy between the lower observed number of satellites in the Local Group and the higher number predicted in numerical simulations \citep{Moore1999,Klypin1999}, also known as the missing satellites problem. 

Up until now, however, the observed data set has been limited. Two recent developments, observation of fainter dwarfs and the discovery of satellites outside the Local Group, for the first time yield data sets that allow for a study of the large-scale statistics of satellites galaxies. 

Large surveys, including the Sloan Digital Sky Survey (SDSS) \citep{Abazajian2009} and PAndAS \citep{McConnachie2008,Martin2009,Richardson2011}, bring observations to even lower surface brightness levels and have revealed a new category of dwarf galaxies: the ultra faint dwarfs \citep{Willman2005,Belokurov2007, Koposov2008}. Currently these large surveys do not cover the entire sky and there may well be ultra faint dwarfs too faint for current instrumentation. \citet{Koposov2008} have calculated a theoretical and complete luminosity function of the Milky Way, based on observations of ultra faint dwarfs and it compensates for the incomplete sky coverage.

Beyond ultra faint dwarfs within our own Local Group, dwarf satellites in other systems are being revealed as well. It is preferable to study satellites in varied environments, as opposed to the Local Group exclusively, even though the Milky Way appears to be typical in terms of the abundance of classical satellites \citep{Strigari2011}. Observations of galaxy clusters have previously yielded the luminosity functions of cluster galaxies \citep[e.g.][]{Abell1977,Jones1980,Binggeli1985} and more recently follow the mass limit down to resolve the satellites of the galaxies themselves. These include M101 \citep{Holmberg1950}, the nearby Virgo Cluster \citep{Sandage1985}, the Fornax Cluster \citep{Ferguson1988}, the Ursa Major Cluster \citep{Trentham2001}, NGC 5846 \citep{Mahdavi2005}, NGC 1407 \citep{Trentham2006}, NGC 5371 \citep{Tully2008}, M81 \citep{Chiboucas2009}, NGC1023 \citep{Trentham2009}, the Coma cluster \citep{Chiboucas2011}, the Antlia cluster \citep{Smith2011}, and other systems using the Sloan Digital Sky Survey \citep{Tollerud2011, Lares2011}. 

These two developments from the observational end have alleviated the traditional missing satellites problem by increasing the number of satellite galaxies. From the theoretical end, simulations that take baryonic physics into account have also helped to close the the discrepancy between the number of observed satellites and predicted dark matter subhalos \citep[e.g.][]{Governato2007, Maccio2009, Okamoto2009, Wadepuhl2010,  Nickerson2011}. In these simulations, a large fraction of subhalos contain only trace quantities of baryons and accordingly may only be observed through gravitational lensing and other gravity-based methods. However, other details still remain to be solved; for example, comparison of the faintest satellites is compromised by the limited resolution of the simulations, and there are discrepancies between the internal kinematics of observed satellites and the structure of the dark matter subhalos they are predicted to inhabit \citep{Boylan2011}.

When studying this new wealth of dwarf galaxies, both observed and simulated, it is also important to consider how the environment, chiefly the properties of the host galaxy, affect the satellites. Previous studies have focussed on brighter and more massive galaxies \citep[e.g.][]{Christlein2000, Nichol2003, Balogh2004}, while we are concerned with how environment affects the newly-observed satellite galaxies in distant systems.

\citet{Trentham2009} studied the spiral-rich group of galaxies around NGC 1023, whose luminosity function has a characteristic magnitude $-25.5 < M^* < -24.0$ and faint-end slope $-1.22<\alpha<-1.14$ \citep{PressSchechter1974,Schecter1976}. They combined their new findings with the host galaxies NGC 1407, NGC 5846, NGC 5353/4, M81, and the Local Group. The galaxies' masses range from a little over $10^{12}$\msun to almost $10^{14}$\msun, and each contain 15 to 250 satellites within the stated credibility limit of $M_R \le -11$. They divided the satellites into giants and dwarfs at $M_R=-17$ and compared the total number of satellites in each category to host mass. For the dwarf satellites they found a tight correlation with host mass:
\beq
N_d \propto M_{host}^{0.91 \pm 0.11};
\label{eqn:ttdwarf}
\eeq
 while giants had a looser correlation: 
 \beq
 N_g \propto M_{host}^{0.67 \pm 0.14}.
 \label{eqn:ttgiant}
 \eeq 
This suggests that the number of dwarf satellites is more accurately  predicted by host mass, while giant satellites are not plentiful enough to provide as reliable statistics.

We combine \citeapos{Trentham2009} data set of dwarf mass to host mass with \citeapos{Koposov2008} theoretically complete luminosity function for the Milky Way to establish a final luminosity function that scales with the mass of a host galaxy. We will verify that our simulations are well-described by this, and also find that the relation remains the same for spiral and elliptical hosts.

In section \S \ref{sec:method} we briefly review the MUGS simulations and halo finding method; in \S \ref{sec:sixteen} we introduce the sixteen host systems studied and detail the luminosity function of each; in \S \ref{sec:data} we compare our satellite luminosity functions to the Press-Schecheter function and the Trentham-Tully relation to justify its use; and our conclusions are \S \ref{sec:conclu}.

\section{Method}
\label{sec:method}

We analyse the subhalos of sixteen galaxies from the McMaster Unbiased Galaxy Survey (MUGS) \citep{Stinson2010}, a sample of M$^*$ galaxies simulated at high resolution. MUGS was run using the SPH code \textsc{gasoline} \citep{Wadsley2004}.  \textsc{gasoline} includes low-temperature metal cooling \citep{Shen2010}, UV background radiation, star formation that models the Kennicutt-Schmidt Law \citep{Kennicutt1998}, and physically-motivated stellar feedback from the ``blastwave model"  \citep{Stinson2006}.  The metal cooling grid is constructed using CLOUDY (version 07.02 \citet{Ferland1998}), assuming ionisation equilibrium. A uniform ultraviolet ionising background, adopted from Haardt \& Madau (in preparation; see \citet{Haardt1996}), is used in order to calculate the metal cooling rates self-consistently. The UV starts to have an effect at $z \sim 9.9$. With these prescriptive elements, \citet{Nickerson2011} found that the resulting model galaxies do not exhibit the missing satellite problem. This results from a combination of early UV heating, ram pressure stripping, tidal stripping and stellar feedback that substantially modify the mass-to-light ratios of the model galaxies' satellites. This is a prediction of the simulations and does not involve any specific tuning in the galaxy model.

We evolve several 50 $h^{-1}$ Mpc volumes of a WMAP3 $\Lambda$CDM universe ($H_{0}$=73 km s$^{-1}$ Mpc$^{-1}$, $\Omega_{m}$=0.24, $\Omega_{\Lambda}$=0.76, $\Omega_{baryon}$=0.04, and $\sigma_{8}$=0.79) \citep{Spergel2007}. From these galaxies we choose a random selection with halo masses between $\approx5\times10^{11}$\msun\ and $\approx2\times10^{12}$\msun\ that did not evolve within 2.7 Mpc of a structure more massive than $5.0\times10^{11}$\msun. The sample is unbiased with regards to angular momentum, merger history, and less massive neighbours and it is desired that the sample will reproduce the observed spread in galaxy properties.  The selected galaxies are then re-simulated with the commonly-used zoom technique, which adds high resolution dark matter and baryons in the region of interest, while maintaining the periferal galaxies at low resolution to provide the appropriate tidal torques. The initial dark matter, gas and star particle masses are $1.1\times 10^{6}$\msun, $2.2 \times 10^{5}$\msun and $6.3 \times 10^{4}$\msun respectively. Each type of particle uses a constant gravitational softening length, 310 pc. A full description of MUGS can be found in \citet{Stinson2010}.  

In order to identify a host galaxy and its subhalos, we use the Amiga Halo Finder (AHF) \citep{Knollmann2009}. AHF is based on the spherical overdensity method to identify density peaks using an adaptive mesh algorithm. AHF cuts out halos (and subhalos) of identified density peaks using isodensity contours.  A simple unbinding procedure is used to determine whether the particles are gravitationally bound to a halo or its subhalos. 

\section{Sixteen Systems of Satellites}
\label{sec:sixteen}
\begin{table*}
\begin{center}
\begin{tabular}{c|ccc|ccc|ccc}
Galaxy & Mass &  Class & R$_{vir}$ & N$_{gas}$ & N$_{star}$ & N$_{dark}$ & Mass$_{sat}$ & N$_{sat}$ & N$_{lume}$ \\
    &   ($10^{11}$\msun)& & (kpc) & $(10^{5})$ & $(10^{6})$ & $(10^{5})$ & ($10^{10}$ M$\sun$)  &&\\ \hline
g7124 & 5.0 & E & 165 & 1.4 & 1.2 & 3.7 & 7.8 & 37 & 9\\
g5664 & 5.7 & S & 173 & 2.0 & 1.1 & 4.3 & 3.3 & 43 & 8\\
g8893 & 6.7 & E &182 & 2.2 & 1.4 & 5.0 & 2.5 & 57 & 10\\
g1536 & 7.5 & S & 190 & 2.7 & 1.4 & 5.7 & 1.5 & 56 & 7\\
g21647 & 8.8 & S & 200 & 3.0 & 1.8 & 6.7 & 14 & 63 & 10\\
g22795 & 9.2 & E & 203 & 3.1 & 1.5 & 7.1 & 3.5 & 76 & 11\\
g22437 & 9.5 & E & 206 & 3.8 & 1.7 & 7.2 & 7.3 & 72 & 9\\
g422 & 11 & S & 218 & 4.0 & 2.3 & 8.5 & 32 & 108 & 16\\
g3021 & 11 & E & 218 & 3.9 & 2.5 & 8.5 & 7.8 & 110 & 17\\
g24334 &12 & S & 221 & 3.7 & 2.5 & 8.9 & 18 & 113 & 26\\
g28547 & 13 & S & 226 & 4.0 & 2.9 & 9.4 & 26 & 107 & 18\\
g4720 & 13 & S & 229 & 5.2 & 2.1 & 10 & 17 & 151 & 27\\
g25271 & 14 & E & 233 & 4.6 & 2.4 & 11 & 3.3 & 96 & 12\\
g15784 &15 & S & 240 & 5.3 & 2.6 & 12 & 11 & 96 & 21\\
g4145 &15 & S & 239 & 5.7 & 2.8 & 1.1 & 19 & 130 & 29\\
g15807 &23 & E & 276 & 8.7 & 4.0 & 17 & 11 & 161 & 33\\
\end{tabular}
\label{tab:gals}
\caption{The attributes of the sixteen MUGS host galaxies at redshift  zero, arranged in order of increasing mass: mass, the classification (\emph{E}lliptical or \emph{S}piral), the virial radius as found by AHF, the number of gas, star and dark particles, the total mass of satellites found by AHF with fifty or more particles at redshift zero, and the number of satellites and the subset of those that are luminous.}
\end{center}
\end{table*}

\begin{figure*}
\begin{center}
\resizebox{\linewidth}{!}{\includegraphics{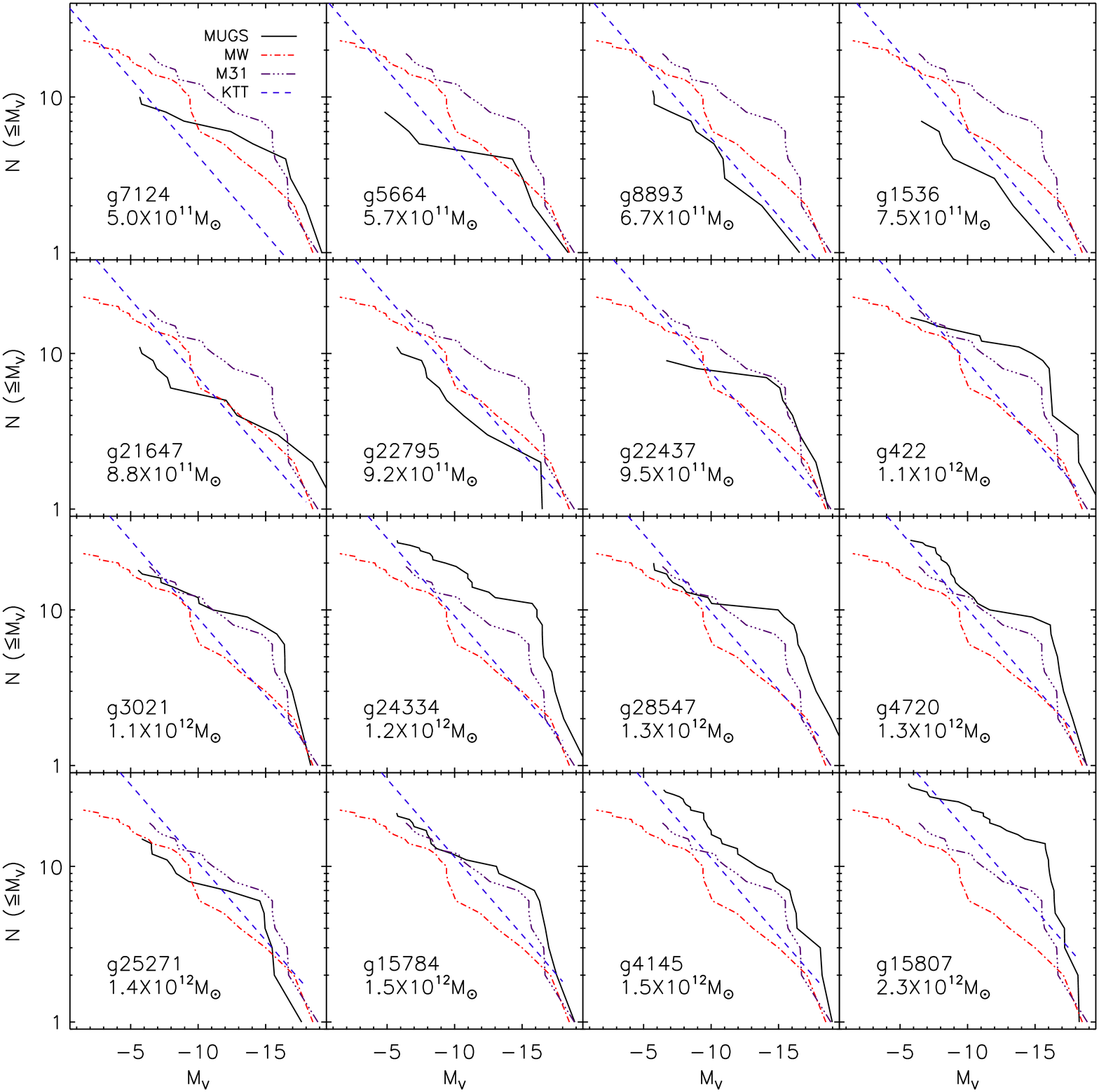}}
\caption{The cumulative V-band luminosity functions of the subhalo populations of the MUGS host galaxies at $z=0$, in solid black,  ordered by host mass, with the host mass given at the bottom. Observational data of the Milky Way \citep{Tollerud2008} is in dashed-dotted red, M31 is in triple-dot-dashed purple \citep{Mateo1998, Brausser2011, Slater2011, Bell2011} and our Equation \ref{eqn:koponorm} (KTT) is in dashed blue taken from \citet{Koposov2009}, normalized to our Equation \ref{eqn:ttdwarf} taken from \citet{Trentham2009}.
. The ultra faint dwarfs as observed in the Milky Way go much fainter than the resolution limit of our simulation.}
\label{fig:cumemag}
\end{center}
\end{figure*}

MUGS consists of sixteen galaxies simulated at high resolution each with identical baryonic physics. Their masses, morphologies, virial radii and the number of gas, star, and dark particles at redshift zero are given in Table \ref{tab:gals}, illustrating the diversity of MUGS. The total mass of satellites in the system found by AHF with fifty or more particles, the number of satellites and the number that are luminous at redshift zero are also included. g8893, g422, g3021, g28547, g4720, g4145, and g15807 are presented here for the first time\footnote{Images of these galaxies can be found at http://mugs.mcmaster.ca/}, while the rest were presented in \citet{Stinson2010}. 

For reference, current estimates of the mass of the Milky Way are around $10^{12}$ \msun \citep{Klypin2002,Bovy2012} and and for M31 are around $5.0\times10^{12}$ \msun \citep{Foreman2010}. Finding these masses is still a very active and ongoing field of research.

Details of how we calculated the luminosity for each subhalo can be found in \citet{Nickerson2011}, using the initial mass functions of  \citet{Kroupa1993} and the luminosity grid from CMD 2.1 \citep{Leitherer1999,Marigo2008}. 

Figure \ref{fig:cumemag} shows the cumulative V-band luminosity function of the subhalo populations of the sixteen host galaxies, ordered by mass, at $z=0$. Also shown is the \citet{Tollerud2008} data for the Milky Way that includes both the classical satellites and the new ultra-faint dwarf galaxies (which are fainter than the resolution of our simulations) and the luminosity function of M31. The classical M31 satellites are taken from \citet{Mateo1998} with the addition of newly-discovered ultra faint dwarfs \citep{Brausser2011, Slater2011, Bell2011}. 

It is believed that the set of ultra faint dwarfs observed in the Local Group is incomplete because of their low surface brightness and incomplete sky coverage of the Sloan Digital Sky Survey. \citet{Koposov2008} provides a theoretical function that would represent a complete set of subhalos with V-band magnitudes from -2 to -11 for the Milky Way given by
\beq
\frac{dN}{dM_V}=10\times10^{0.1(M_V+5)}
\label{eqn:kopo}
\eeq
where $N$ is the number of satellites that have a magnitude of $M_V$ or brighter. However, in order to look at the luminosities for systems of satellites in non-Milky Way-like hosts we need to readjust the normalization, i.e. the total number of satellites expected in different sized galaxies. For this we use the Trentham-Tully relation \citep{Trentham2009}:
\beq
\mathrm{log}N_d=-10.2(\pm1.4)+0.91(\pm0.11)\mathrm{log}M_{host}
\label{eqn:ttdw}
\eeq
where $N_d$ is the total number of dwarf galaxies with R-band magnitudes between -11 and -17 for host of mass $M_{host}$ in $\Msun$. We only use the dwarf satellites for this and not the giants, for reasons that are detailed in \S \ref{sec:data}. Some speculation is involved here because \citet{Koposov2008} and \citet{Tollerud2008} have differing magnitude limits, \citet{Tollerud2008} being limited by low surface brightness, which is why we will need to test this in \S \ref{sec:data} and ensure that it works. Because the range in the Trentham-Tully relation is given in the R-band, while the V-band is used for the Koposov function, we need to convert the range. From our sample of 280 luminous satellites in MUGS, we find that $V-R=0.51\pm0.01$. The final expression for the luminosity function as a function of the host mass and V-band magnitude is:
\beq
\frac{dN}{dM_V}=3.5M_{host}^{0.91}\times10^{0.1M_V-10.2}
\label{eqn:koponorm}
\eeq
and this is also shown in Figure \ref{fig:cumemag} alongside the luminosity functions of the MUGS simulations.

The Milky Way and M31 luminosity functions have different shapes. The Milky Way remains as a steady power law for high luminosity satellites. M31 has a knee, showing a higher number of high luminosity satellites. Correspondingly, a few of our galaxies also match the M31 shape better than the Milky Way's, notably g422, g3021 and g15784. On the other hand, g8893, g22795, and g4145 match the shape of the Milky Way luminosity function, as given by \citet{Tollerud2008} closely. However, most other galaxies display an excess knee in high luminosity satellites compared to the Milky Way. We will now examine the total number itself.

Of the galaxies that are less massive than the Milky Way and go up to its mass, ranging from $\sim 50 \%$ to $95 \%$ of its mass, g7124 and g8893 have a similar cumulative number of satellites compared to the normalized power law Equation \ref{eqn:koponorm}. g5664, g1536, g21647, g22795, and g22437 are relatively satellite-poor, though are still within an order of magnitude of what is predicted. 

The Milky Way-like-in-size galaxies and slightly more massive (g422, g3021, g24334, g28547, g41720, g25271) have very similar cumulative luminosity functions to the \citet{Tollerud2008} function for their brightness limit, or are near to the Milky Way's function. The MUGS curves do not appear to dramatically change shape as they approach the resolution limit, which suggests that the total number of low luminosity satellites is relatively robust.

Of the heavier galaxies that exceed the Milky Way's mass, g15784 is the only one of these that meets M31's satellite count and g4145 and g15807 match Equation \ref{eqn:koponorm}. Our high luminosity satellites are probably more luminous than is realistic, but they are within the resolved range and should still contribute to the total cumulative number.

The importance of having a sample of galaxies in different environments is demonstrated by the variance in luminosity functions across galaxies of similar mass. Overall, in terms of cumulative number, and within the resolution limit, our galaxies do not suffer from the order of magnitude missing satellites problem as is evident from Equation \ref{eqn:koponorm}. In spite of the differing environments, a trend still holds to predict the number of satellites based purely on the host's mass. Only the satellites below the Milky Way's mass have slightly fewer satellites than expected, while every Milky Way-mass and more massive host meet this trend. This might be an effect of small scale statistics for the lower mass hosts and with higher resolution we expect they too should follow the trend closely. 

\section{The Robustness of Dwarf Satellites over Giant Satellites}
\label{sec:data}

\begin{figure}
\resizebox{\linewidth}{!}{\includegraphics{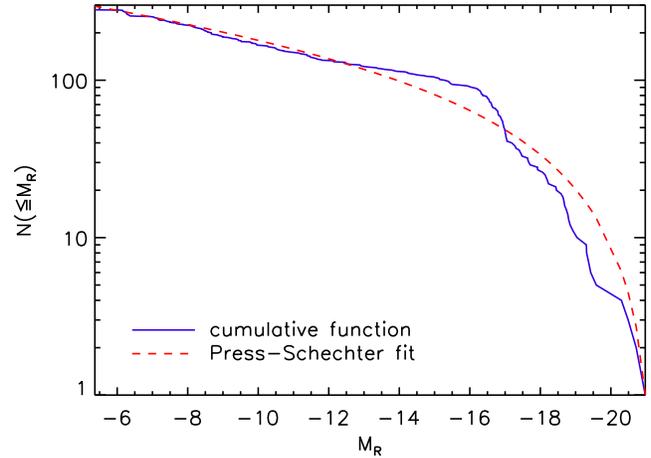}}
\caption{The cumulative R-band luminosity function for all satellites of the MUGS hosts in solid blue. The Press-Schechter functional fit is in dashed red.}
\label{fig:mrcume}
\end{figure}

Figure \ref{fig:mrcume} shows the cumulative luminosity function for all the satellites of the sixteen MUGS galaxies. It includes a fit to the Press-Schechter function \citep{PressSchechter1974}, adjusted to an intercept such that the brightest satellite has a cumulative number of 1. In cumulative form \citep{Schecter1976} the function is,
\beq
N_e = \Gamma(\alpha+1,L/L^*)n^*
\label{eqn:ps}
\eeq
where their $N_e$ is the number of galaxies expected to have a higher luminosity than $L$, $\alpha$ is the faint-end slope, $L^*$ is the characteristic luminosity at which the faint end begins, $n^*$ is the richness parameter, and $\Gamma$ is the incomplete upper gamma function. 

Most of our galaxies display an excess of high-luminosity satellites, making the fit to the Press-Schechter function somewhat problematic. We were nevertheless able to obtain a fit that brackets the high-luminosity bend: $\alpha+1=-5.48 \pm 0.67 \times 10^{-2}$ and $M^*_r =-21.3 \pm 0.2$. It is already clear from Figure \ref{fig:mrcume} that our dwarf satellites fit the Press-Schecter functional form much better than the giant satellites.

\begin{figure}
\resizebox{\linewidth}{!}{\includegraphics{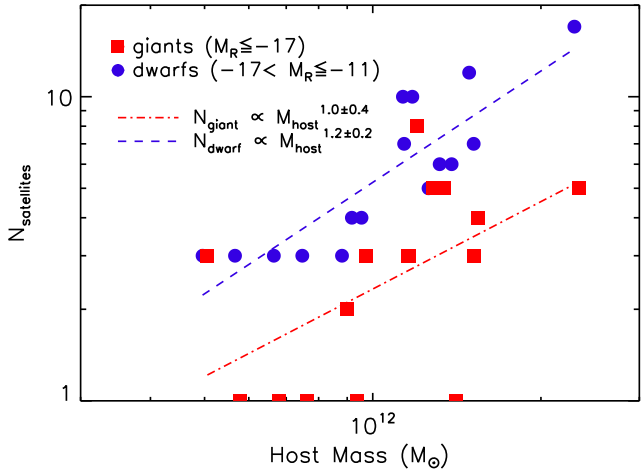}}
\caption{The number of satellites versus host mass. The satellites are divided at R-band magnitude -17 between giants (red squares) and dwarfs (blue circles). The power law fit for the giants is in dotted-dashed red, and the fit for the dwarfs is in dashed blue. The giants are offset to the right slightly so that if they overlap in number with the dwarfs for a particular galaxy both points can still be seen.}
\label{fig:mhost}
\end{figure}

Figure \ref{fig:mhost} shows the number of luminous satellites versus the mass of the host galaxy. As with \citet{Trentham2009} we do not consider satellite galaxies fainter than $M_R= -11$, and we split the dwarfs from the giant satellites at $M_R= -17$. The power law fits for this relation are shown in Figure \ref{fig:mhost}. We obtain for dwarfs,
\beq
N_d \propto M_{host}^{1.2 \pm 0.2}
\label{eqn:nmdwarf}
\eeq
 and for giants, 
 \beq
 N_g \propto M_{host}^{1.0 \pm 0.4}.
 \label{eqn:nmgiant}
 \eeq 
Within the errors, our trends for the dwarfs and giants fall within the Trentham-Tully relations given in Equations \ref{eqn:ttdwarf} and \ref{eqn:ttgiant}. We find, just as they do, that the giant satellites have more scatter when fit to a power law than the dwarf satellites.

\begin{figure}
\resizebox{\linewidth}{!}{\includegraphics{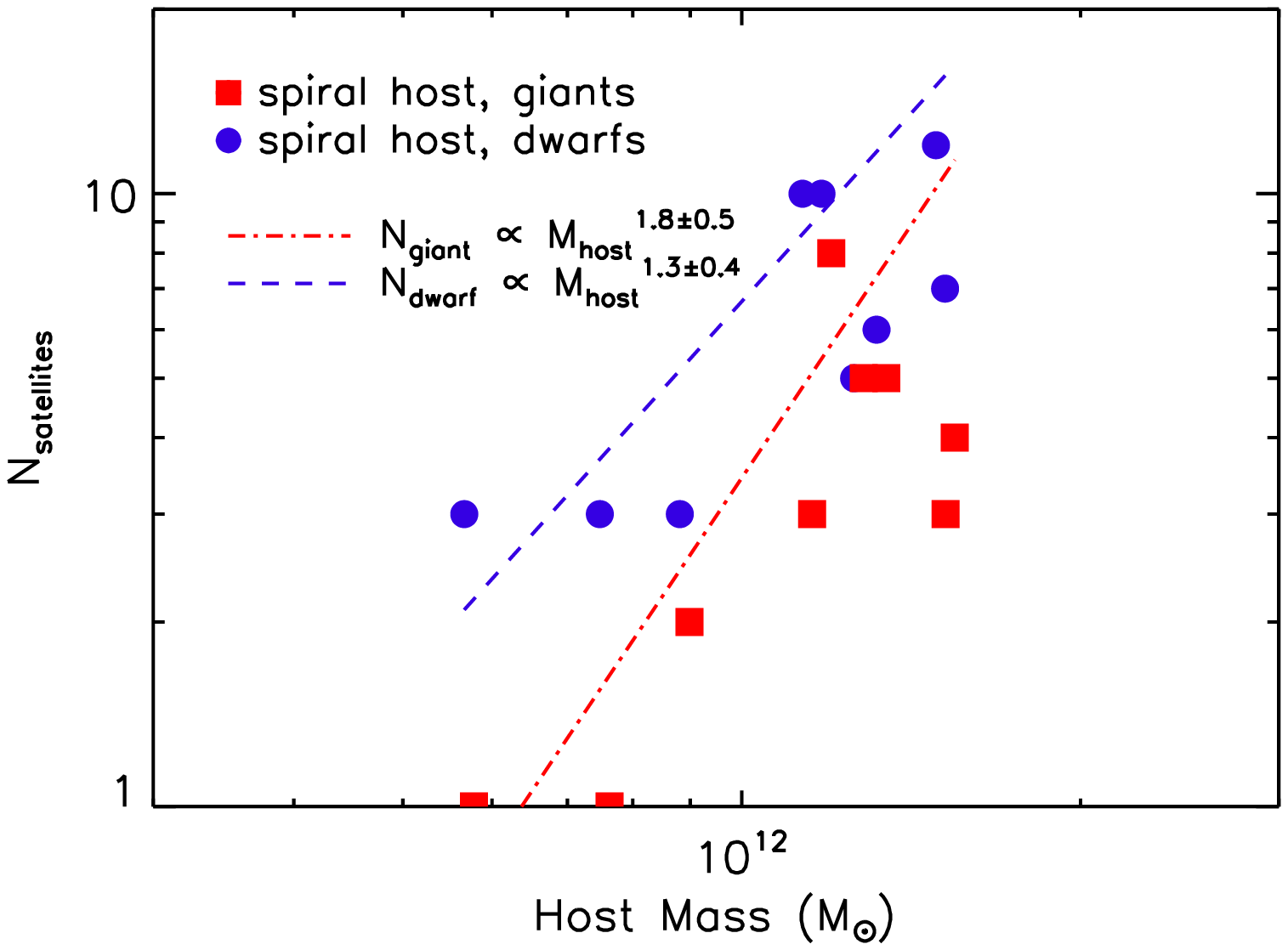}}
\caption{The number of satellites versus host mass for spiral host galaxies only. The satellites are divided at R-band magnitude -17 between giants (red squares) and dwarfs (blue circles). The power law fit for the giants is in dotted-dashed red, and the fit for the dwarfs is in dashed blue. The giants are offset to the right slightly so that if they overlap in number with the dwarfs for a particular galaxy both points can still be seen.}
\label{fig:mhostspiral}
\end{figure}

\begin{figure}
\resizebox{\linewidth}{!}{\includegraphics{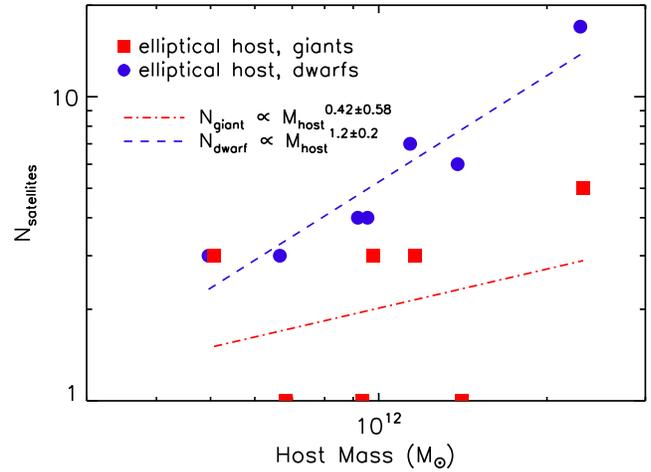}}
\caption{The number of satellites versus host mass for elliptical host galaxies only. The satellites are divided at R-band magnitude -17 between giants (red squares) and dwarfs (blue circles). The power law fit for the giants is in dotted-dashed red, and the fit for the dwarfs is in dashed blue. The giants are offset to the right slightly so that if they overlap in number with the dwarfs for a particular galaxy both points can still be seen.}
\label{fig:mhostellip}
\end{figure}

We investigated splitting the hosts by morphology and exploring the difference in the power law between spiral and elliptical hosts in Figures \ref{fig:mhostspiral} and \ref{fig:mhostellip}. While the power law for the number of dwarf satellites ($1.3 \pm 0.4$ for spirals and $1.2 \pm 0.2$ for ellipticals) remains nearly the same, the power for the giant satellites changes drastically between the two morphologies ($1.8 \pm 0.5$ for spirals and  $0.42 \pm 0.58$ ellipticals). This suggests further that the small-number statistics of giants are more affected by environment, while the number of dwarf satellites is robust against morphology type. 

Dwarf satellites are more useful for finding a universal scaling relation for the luminosity function of satellites according to host mass as in Equation \ref{eqn:koponorm}. These observations verify the trend shown in our simulations, in that the mass of the host predicts the number of dwarf satellites, and hence within resolution our simulation does not suffer from the missing satellites problem. This is an important loadstone for studying luminosity functions and satellite systems outside of the Local Group.

\section{Conclusions}
\label{sec:conclu}
We compare the satellite populations in simulations of sixteen high resolution galaxies with observations to see how satellite populations vary as a function of host mass. Our sample of hosts contains a wide range of masses and morphologies and are therefore interesting to compare to the new studies and statistics of satellite galaxies in diverse systems beyond the Local Group. Just as there are observations of diverse galaxies, there needs to be diverse simulations run with the same physics that contain more than Milky Way-like galaxies. 

The luminosity functions of our galaxies scale well with observations. It shows that with baryon physics, our simulations no longer suffer from the order of magnitude missing satellites problem. For host galaxies as massive as the Milky Way and more massive, the scaling fits almost exactly. Though our hosts with masses lower than the Milky Way do not quite have as many satellites as predicted, they were still only short by three or four satellites, and probably with higher resolution would suffer less from small number statistics.

\citeapos{Trentham2009} study concluded that dwarf galaxy populations ($-17< M_R < -11$) correlate to their host halo's mass, while giants ($  M_R < -17$) do not. They explored the space of host masses from $~ 10^{12}$ to $~ 10^{14}$ \msun\ and found the relation $N_{dwarf} \propto M_{host}^{0.91 \pm 0.11}$, where mass is given in $\Msun$. We find that that the Trentham-Tully relation describes our simulated satellite systems even when we split our hosts between spirals and ellipticals. The number of dwarf satellites a host has is largely dependent on the host mass and does not change with morphology type, whereas the number of giant satellites might be affected by many factors. There were not enough giants on which to found solid statistics.

When all satellite galaxies from all sixteen MUGS simulations are stacked together, their luminosity function follows a Press-Schechter function with faint-end slope $\alpha+1=-5.48 \pm 0.67 \times 10^{-2}$ and characteristic luminosity $M^*_r =-21.3 \pm 0.2$.

\citet{Koposov2008} provided a luminosity function for the Milky Way that takes into account the incomplete sky coverage of large surveys and dwarfs too faint for detection. We use the knowledge of dwarf satellites' robustness to adjust the Milky Way's function in order to incorporate dependence on host mass, as in the Trentham-Tully relation for dwarfs: $dN/dM_V=3.5M_{host}^{0.91}\times10^{0.1M_V-10.2}$, where mass is given in $\Msun$. The MUGS hosts compare much more favourably to this new, scaled luminosity function than they do to the Milky Way's or M31's luminosity functions alone.

\section*{Acknowledgements}
We thank SHARCNET for providing supercomputer time without which the MUGS galaxies would not have been possible, and NSERC for funding. HMPC thanks the Canadian Institute for Advanced Research for support.

\bibliographystyle{mn2e}
\bibliography{references}

\clearpage

 \end{document}